\documentclass[preprint,superscriptaddress,groupedaddress]{revtex4}
\usepackage[latin9]{inputenc}
\usepackage{amsbsy}
\usepackage{amstext}
\usepackage{amssymb}
\usepackage{graphicx}
\usepackage{esint}

\makeatletter
\@ifundefined{textcolor}{}
{%
 \definecolor{BLACK}{gray}{0}
 \definecolor{WHITE}{gray}{1}
 \definecolor{RED}{rgb}{1,0,0}
 \definecolor{GREEN}{rgb}{0,1,0}
 \definecolor{BLUE}{rgb}{0,0,1}
 \definecolor{CYAN}{cmyk}{1,0,0,0}
 \definecolor{MAGENTA}{cmyk}{0,1,0,0}
 \definecolor{YELLOW}{cmyk}{0,0,1,0}
}

\usepackage{dcolumn}
\usepackage{bm}

\makeatother

\begin{document}


\title{Giant perpendicular magnetic anisotropy in Fe/III-V nitride thin films}

\author{Jie-Xiang Yu}

\affiliation{Department of Physics and Materials Science Program, University of New Hampshire, Durham, New Hampshire 03824, USA}

\author{Jiadong Zang}

\affiliation{Department of Physics and Materials Science Program, University of New Hampshire, Durham, New Hampshire 03824, USA}

\date{\today}
\begin{abstract}
Enabling large perpendicular magnetic anisotropy (PMA) in transition metal thin films is a pathway towards intriguing physics of nanomagnetism and broad spintronics applications. 
After decades of material searches, the energy scale of PMA of transition metal thin films, unfortunately, remains only about 1 meV.
It has become a major bottleneck towards the development of ultradense storage and memory devices.
Here, we discovered unprecedented PMA in the Fe thin films growth on $(000\bar{1})$ N-terminated surface of III-V nitrides from first-principles calculations. PMA ranges from 24.1 meV/u.c in Fe/BN to 53.7 meV/u.c. in Fe/InN. Symmetry-protected degeneracy between $x^2-y^2$ and $xy$ orbitals and its lift by the spin-orbit coupling play a dominant role. As a consequence, PMA in Fe/III-V nitride thin film is predominated by the first order perturbation of the spin-orbit coupling, instead of second order in conventional transition metal/oxide thin films. 
This game-changing scenario would also open a new field of magnetism on transition metal/nitride interfaces.
\end{abstract}
\maketitle

\subsection*{INTRODUCTION}
Magnetic anisotropy is a relativistic effect originating from the spin-oribt coupling (SOC).
Perpendicular magnetic anisotropy (PMA) in magnetic thin films has led to rich physics and become a key driving force in the development of magnetic random-access memory (MRAM) devices\cite{Chappert2007, Brataas2012, Dieny2017}. 
Establishment of PMA in nanostructures and nanopatterned magnetic multilayers paves a new avenue towards nanomagnetism, in which fascinating physics such as spin Hall switching and skyrmions are blooming\cite{Miron2011, Liu2012a, Liu2012b, Jiang2017}. 
As the strength of the SOC is a quartic function of the atomic number, it is not surprising to have large magnetic anisotropy in heavy metals such as rare earth materials, which are commonly used for permanent magnets \cite{Sagawa1984,Sugimoto2011}.
It is, however, a challenge to enable large anisotropy, especially PMA, in commonly used $3d$ transition metals such as Fe thin films.

Actually strength of PMA is determined by the energy correction from the SOC, which couples the orbital angular momentum $\bf{L}$ to the spin momentum $\bf{S}$ via $H_{so}=\lambda \bf{L}\cdot \bf{S}$.
In the single iron atom limit, 6 valence electrons of Fe in $3d$ shell could ideally have a total spin of $S_z=2$ and angular momentum of $L_z=2$.
The atomic limit of the SOC energy $\lambda \bf{L}\cdot \bf{S}$ is thus 75 meV given by the SOC coefficient $\lambda\approx19$ meV \cite{SOC_1962}.
However, in all existing discussions of Fe-based thin film on MgO substrates, such as Fe/MgO\cite{Wang1993, Maruyama_2009, Yang2011, Koziol-Rachwal2014, Ibrahim2016} and CoFeB/MgO-based system \cite{Ikeda2010,  Worledge2011, Liu2014}, the size of PMA is only 1 meV, far below the atomic limit.
This leaves a vast window to escalate PMA in transition metal thin films unexploited.

Under crystal field, five $d$-orbitals are superposed and form $xy$, $yz$, $xz$, $x^2-y^2$, and $3z^2-r^2$ orbitals as the eigenstates.
All of the new orbitals have zero$L_z$ due to the time reversal symmetry.
If these orbitals are nondegenerate, the first order energy correction from SOC vanishes, leaving the second order perturbation as the dominant contribution \cite{Wang1993, WuRQ1999}. 
This is the scenario in most thin film systems 
\cite{Wang1993, Maruyama_2009, Yang2011, Koziol-Rachwal2014, Ibrahim2016, Ikeda2010, Worledge2011, Liu2014, Yang2017}.  
The energy scale of PMA is then $\lambda^2/\Delta$, where $\Delta$ is the band width of the state crossing the Fermi level. 
For a typical $3d$ magnetic element, $\Delta \sim 1$ eV and $\lambda \sim 0.03$ eV \cite{SOC_1962}.
It is thus not surprising to achieve 1 meV PMA in most $3d$ magnet thin films.

\begin{figure*}
\includegraphics[scale=0.50]{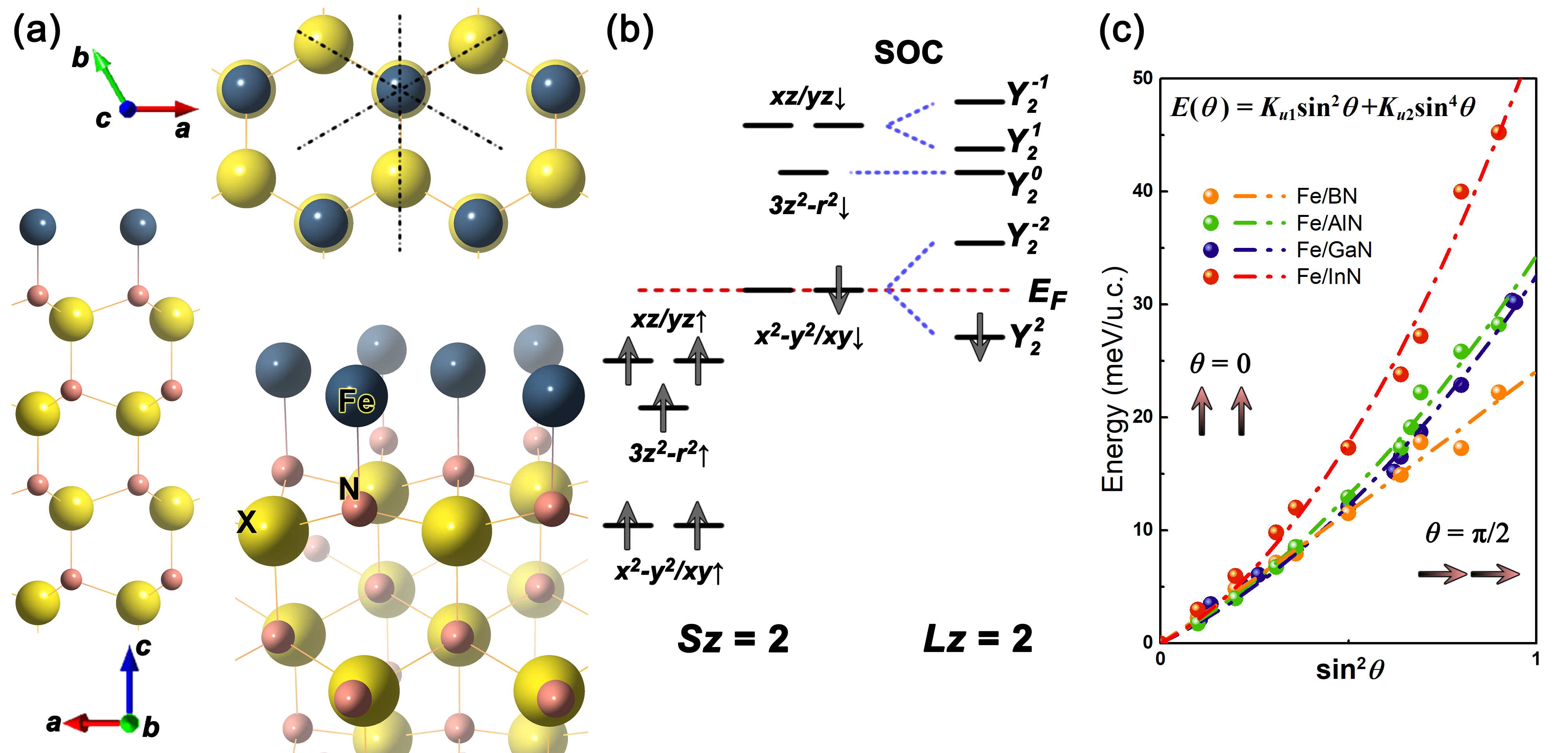}
\caption{Perpendicular magnetic anisotropy of Fe(1ML)/III-V nitrides. 
(a) The surface structures of one monolayer (1ML) Fe deposited on  $(000\bar{1})$ N-terminated surface of III-V nitride XN (X = B, Al, Ga and In) substrate in top, side and perspective view, respectively. 
In top view, three dash-dotted lines indicate three mirror planes of the $C_{3v}$ point group on Fe.  
(b) Crystal field diagram for Fe($3d$)'s spin majority (spin up) and minority (spin down) channels. The spin minority channel is further split by SOC. $E_F$ refers to the Fermi level. (c) The relative total energy per unit cell (u.c.) of Fe(1mL)/III-V nitride thin film as a function of $\sin^2\theta$, where $\theta$ denotes the angle between magnetization orientation and the $z$ direction. Relative energy at $\sin^2\theta=1$ gives the value of PMA. The dotted curves are their fit to $K_{u1}\sin^2\theta+K_{u2} \sin^4\theta$.}
\label{fig:intro} 
\end{figure*}

To escalate the PMA, one thus would like to find a regime the first order perturbation of SOC is dominant. 
In this regime, PMA is proportional to $\lambda$ instead and has the chance to approach the atomic limit of SOC energy. 
It occurs when partially filled degenerate orbitals exist around the Fermi level. 
%
A successful example has already been demonstrated in a single adatom\cite{Rau2014, Khajetoorians2014, Ou2015, Blonski2010} or dimer \cite{Donati2013, Hu2014} deposited on specific substrates.
%
However, once a thin film is formed, PMA in these systems is greatly reduced and brought back to the second order perturbation scheme. 
Here we report giant PMA in Fe ultrathin films grown on the wurtzite $(000\bar{1})$ N-terminated surface of III-V nitrides XN, where X = B, Al, Ga and In [Fig.\ref{fig:intro}(a)].
First order perturbation of SOC is exactly the mechanism responsible, and the atomic energy limit is approached.
%

\subsection*{RESULTS}
As a central result of this work, total energies with different magnetization directions of 1ML Fe on BN, AlN, GaN and InN, respectively, were obtained. 
The relative magnetoanisotropy energy as a function of $\sin^2\theta$ is shown in Fig.\ref{fig:intro}, where $\theta$ is the angle between the magnetization and $z$ direction. 
At small $\theta$, a linear relation between magnetoanisotropy energy and $\sin^2\theta $ is observed. 
It is consistent with well-adopted descriptions of PMA in most thin film systems \cite{Laan1998, Dieny2017}.
The lowest energy lies at $\theta=0$ so that the uniaxial anisotropy along the $z$ direction is clearly identified.
When $\theta$ approaches to angle $\pi/2$, a clear deviation from $\sin^2\theta$ is present, and high order anisotropy $K_{u2} \sin^4\theta$ contributes to PMA as well. 
Therefore, we fit the magnetoanisotropy energy to $K_{u1} \sin^4\theta + K_{u2} \sin^4\theta$\cite{WuRQ1999, Wang1996}.
PMA, the energy difference between perpendicular magnetization and in-plane magnetization, is thus given by $K_{u1}+K_{u2}$.
Fitting parameters for Fe/BN, Fe/AlN, Fe/GaN and Fe/InN are listed in Table \ref{table:pma}.
No significant changes of $K_{u1}$ are found while the fourth order term grows significantly from BN to InN. 
Except for Fe/BN, the contribution of $K_{u2} \sin^4\theta$ to PMA is considerable and is even about twice the second order contribution in Fe/InN.
The resulting PMA values range from 24.1 to 53.7 meV/u.c.. 
They are all over one order of magnitude larger than the PMA of Fe/MgO, which is around $1 \sim 2$ meV/u.c. \cite{Wang1993, Maruyama_2009, Yang2011, Koziol-Rachwal2014, Ibrahim2016}.
Particularly, PMA of Fe(1ML)/InN as 53.7 meV/u.c. approaches the atomic limit of the SOC energy of 75 meV for an isolated Fe.
The PMA values for other three are also on the same order as that limit.

\begin{table}
\caption{PMA and relevant magnetic properties for each Fe(1ML)/III-V nitride. The value $K_{u1}$, $K_{u2}$, total PMA in units meV/u.c. and mJ/m$^{2}$, spin moments $m_s$, orbital moments $m_l$, and occupation number $o_{2}^{2}$ of $Y_{2}^{2}$-dominated state in spin minority are listed.}
\begin{tabular}{ c | c c c c}
\hline \hline
  & Fe/BN & Fe/AlN & Fe/GaN & Fe/InN \\
\hline
 $K_{u1}$ (meV/u.c.) & 22.5 & 18.1 & 16.2 & 17.5 \\
 $K_{u2}$ (meV/u.c.) & 1.6 & 16.2 & 16.4 & 36.2 \\
 PMA (meV/u.c.) & 24.1 & 34.3 & 32.5 & 53.7 \\
 PMA (mJ/m$^{2}$) & 59.1 & 56.6 & 51.3 & 71.9 \\
\hline
 $m_s$ ($\mu_B$) & 3.56 & 3.83 & 3.84 & 3.84 \\
 $m_l$ ($\mu_B$) & 0.91 & 1.44 & 1.54 & 1.51 \\
 $o_{2}^{2}$ & 0.724 & 0.854 & 0.904 & 0.930 \\
\hline \hline
\end{tabular}
\label{table:pma}
\end{table}

\begin{table}
\caption{Bader charges on Fe and the top N atom. Bulk refers to the bulk Fe and GaN respectively; surface refers to the clean GaN $(000\bar{1})$ N-terminated surface; interface refers to 1ML Fe on GaN substrate; the last column gives the results with SOC included.}
\begin{tabular}{ c | c c c c}
\hline \hline
  Bader charge & Bulk & Surface & Interface & Interface SOC  \\
\hline
 Fe & 8.00 & - & 7.61 & 7.60 \\
 N in GaN & 6.52 & 6.15 & 6.54 & 6.54 \\
\hline \hline
\end{tabular}
\label{table:bader}
\end{table}

\subsection*{DISCUSSION}
\begin{figure*}
\includegraphics[scale=0.25]{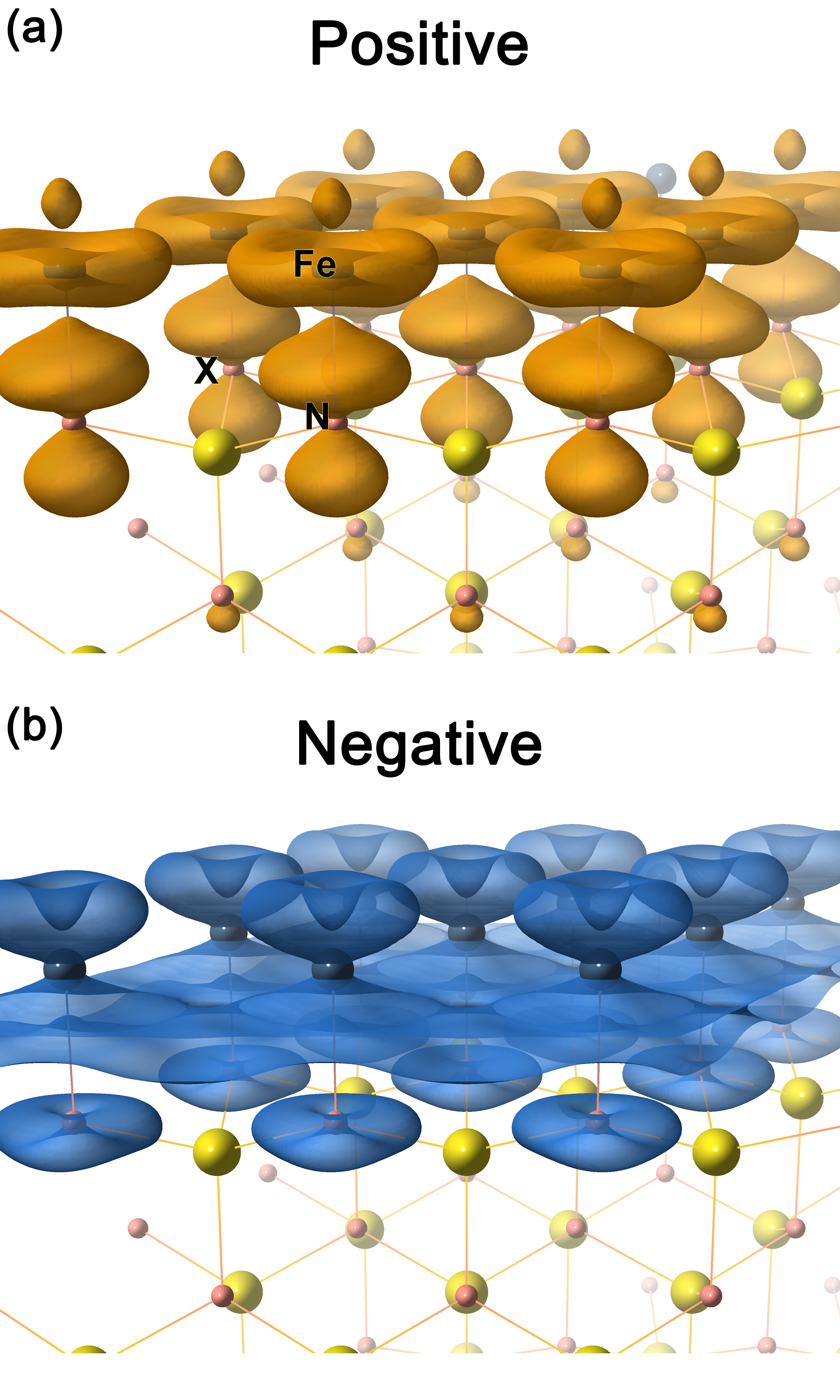}
\includegraphics[scale=0.25]{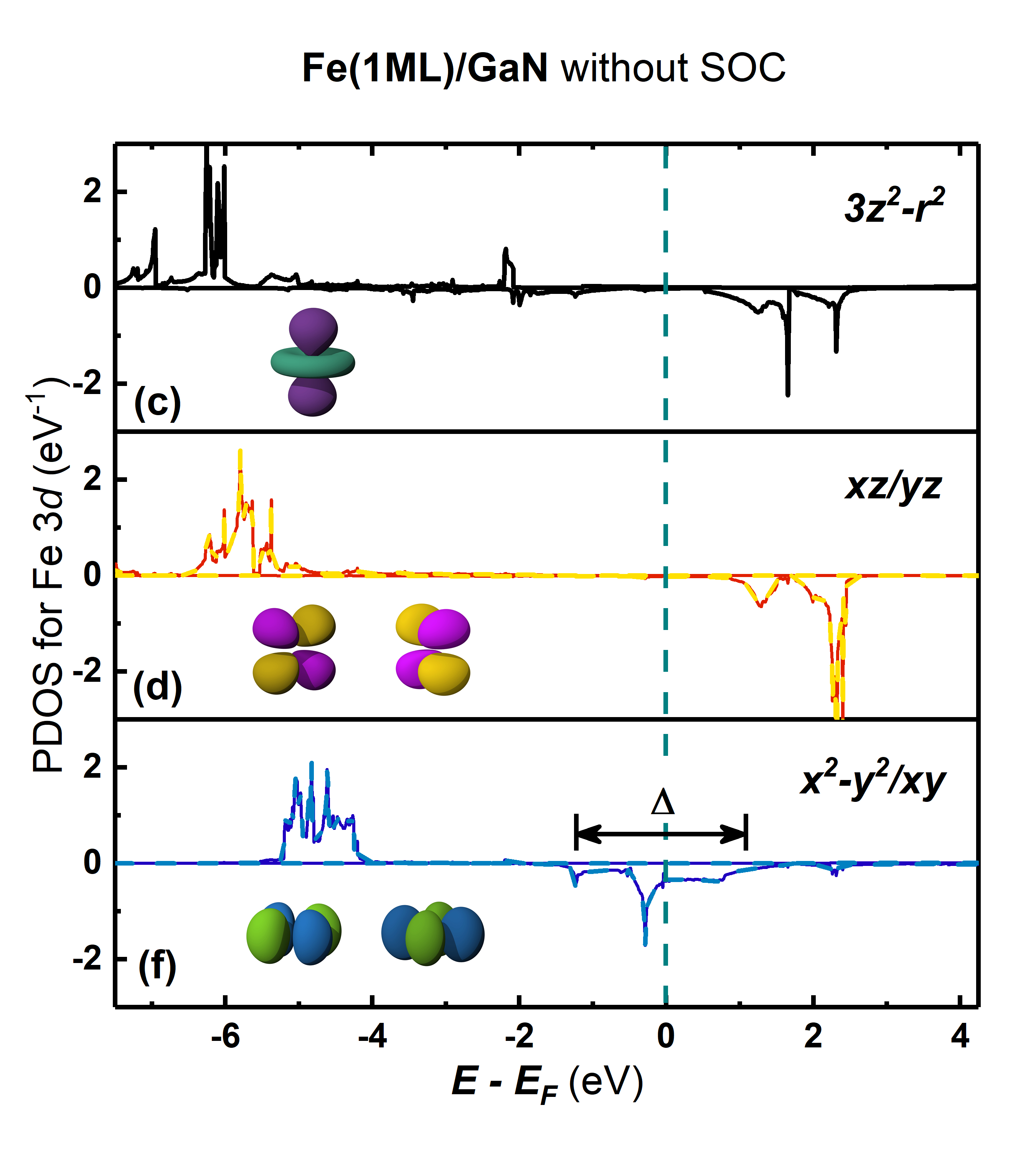}
\includegraphics[scale=0.25]{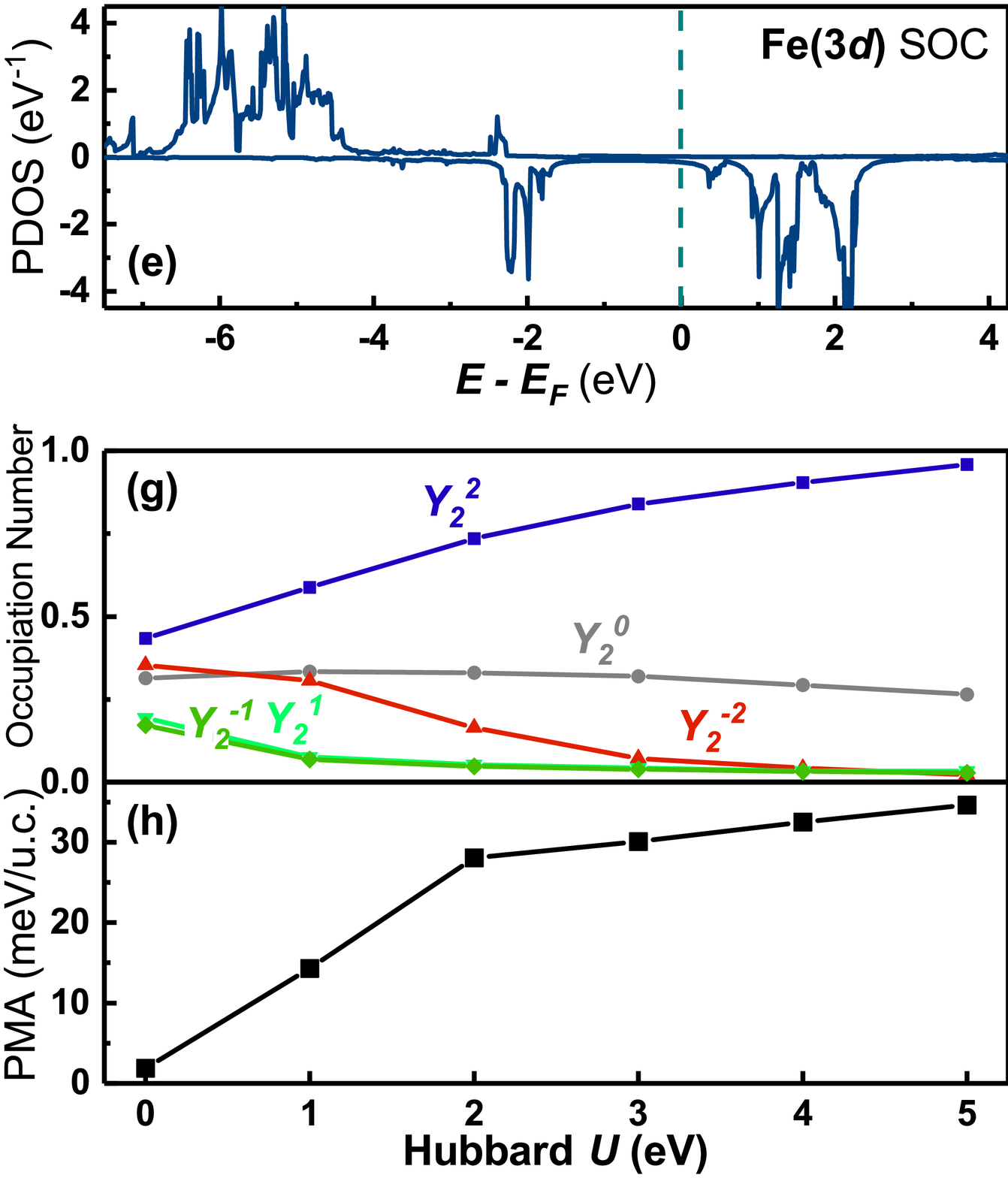}
\caption{Charge distribution and electronic structure of Fe(3d) orbitals in Fe(1ML)/GaN. 
(a) the positive and (b) negative part of the charge difference between the total charge density of Fe(1ML)/GaN and the sum of charge densities of a suspended 1ML Fe and pure GaN supercell. 
Charge deficiency in $xz/yz$-like orbitals shown in (a) is transferred to $x^2-y^2/xy$-like orbitals shown in (b).
(c)-(f) Orbital-resolved projected density-of-state (PDOS) of $3z^2-r^2$, $xz/yz$ and $x^2-y^2/xy$ orbitals, respectively, in the absence of SOC. 
The band width $\Delta$ of $x^2-y^2/xy$ is labeled in (f).
(e) PDOS for Fe($3d$) orbitals with SOC included.
Positive and negative values of PDOS refer to spin majority and spin minority channels respectively, and the Fermi level is set to zero.
(g) The occupation number of each Fe($3d$) orbital and (h) PMA as a function of the Hubbard $U$ when SOC is included. }
\label{fig:orbital} 
\end{figure*}

The giant PMA in Fe/III-V nitrides is considerably beyond the energy scale of the second order perturbation of SOC..
In order to understand the origin of this giant PMA, the electronic structure of Fe($3d$) orbitals were studied.
Without loss of generality, the Fe(1ML)/GaN system was analyzed in detail below. 
Fig.\ref{fig:orbital}(a)(b) displays the difference between total charge density of our Fe(1ML)/GaN system, and the sum of charge densities of a suspended 1ML Fe and a pure GaN supercell. 
Electron density is reduced in blue contours while increased in yellow ones. 
Thus, charge transfer occurs from blue contour to the yellow contour during the formation of the Fe-GaN interface. 
The yellow contour indicates the formation of strongly polarized Fe-N bonds and the enhancement of in-plane $x^2-y^2$ and $xy$ orbitals. 
From blue contours, a significant reduction of Fe's itinerant and $xz/yz$ electrons are witnessed 
The reduction of itinerant electrons is reasonable, itinerant electrons because Fe electrons saturate the dangling bonds from N atoms on N-terminated surface, so that Fe atoms lose electrons and become cations. 
The ionic behaviors of Fe are doubly confirmed by the Bader charge \cite{Bader_01, Bader_02, Bader_03} results (Table \ref{table:bader}), of which the difference corresponds to the charge increasing/decreasing on one atom. 
About 0.4 $e^-$ electrons per Fe atom are transferred to N atoms on the interface. 
These interfacial N atoms thus have almost the same number of valence electron as that in bulk GaN. 
In addition, there is no additional inter-atomic charge transfer when SOC is included, shown on the last column in Table \ref{table:bader}.

To explain the charge transfer from $xz/yz$ to $x^2-y^2/xy$ orbitals and thereby identify valence states of Fe cations, we investigated the crystal field and orbital-resolved projected density-of-state (PDOS) of Fe($3d$) without SOC first.
As shown in Fig.\ref{fig:orbital}(c)-(f), all five $d$ orbitals in the spin majority channel are far below the Fermi level and fully filled. 
Rich physics is present in the spin minority channel. 
Three orbitals predominated by $xz$, $yz$ and $3z^2-r^2$ respectively are far above the Fermi surface and almost unoccupied. 
Double degenerate orbitals, labelled as $e$ orbitals, predominated by $x^2-y^2/xy$ orbitals are low lying [Fig. \ref{fig:orbital}(f)]. 
Importantly, they are crossing the Fermi level and thus partially occupied. 
In Fig.\ref{fig:orbital}(d),  the two fold degeneracy of the $xz/yz$-predominated orbitals, labelled as $e'$ orbitals, is explicitly shown. 
Double degeneracies of $e$ and $e'$ orbitals are protected by the two-dimensional irreducible representation $E$ of $C_{3v}$ point group of the crystal field around each Fe cation. 
In reality, overlap between $xz/yz$ and $x^2-y^2/xy$ orbitals is present, but small. 
According to the density matrix of Fe($3d$) orbitals and the corresponding occupation number, $e$ states are mixed with 3\% $xz/yz$ orbitals, and $e'$ orbitals contain 3\% $x^2-y^2/xy$ components. 
In spin minority channel, $e$ states are almost half filled, with an occupation number of 0.457, while occupation number of $e'$ states is only 0.035. 
As a comparison, in suspended 1ML Fe, the occupation number of $x^2-y^2/xy$ is 0.095, and that of $xz/yz$ is 0.564. 
Therefore $x^2-y^2/xy$ orbitals have escalated occupation once Fe is deposited on GaN. 
It is consistent with the charge density contours discussed earlier. 

Once SOC is included, one can expect the lift of degeneracy between $x^2-y^2$ and $xy$ orbitals due to nonzero off-diagonal matrix elements. 
It is confirmed by the PDOS shown in Fig.\ref{fig:orbital}(e), where almost no states in the spin minority channel are found near the Fermi level. Degenerate $e$ states are split and a large splitting about 3.0 eV is present. 
According to the density matrix, the occupation number of the lower splitting state is 0.904, and the corresponding eigenstate is  
$\sqrt{\alpha}(i\mid xy \rangle+\mid x^2-y^2 \rangle)+\sqrt{\beta}(i\mid xz \rangle-\mid yz \rangle)$
,or equivalently
\begin{equation}
\sqrt{2\alpha}\mid Y_{2}^{2}\rangle+\sqrt{2\beta}\mid Y_{2}^{1}\rangle.
\end{equation}
In terms of spherical harmonics, this state is quite close to $Y_{2}^{2}$, which has a nonzero expectation value of the SOC energy.
Similarly, the higher splitting state is close to $Y_{2}^{-2}$ but hybrid with $e'$ states.
Occupation numbers of those three states are 0.042, 0.033 and 0.032 respectively; that is, almost empty. 
The $3z^2-r^2$, {\it i.e.} $Y_{2}^{0}$, -dominated state is insensitive to SOC, and its occupation number is 0.293. 
Therefore, the net orbital magnetic moment on Fe($3d$) along $z$ direction is 1.54 $\mu_{B}$, consistent with $L_z=2$ due to the splitting into $Y_{2}^{2}$ and $Y_{2}^{-2}$ in $e$ states near the Fermi level. 
The spin moment in the unit cell from this self-consistent calculation is 3.84 $\mu_B$. 
It is quite consistent with $S = 2$, the high spin configuration of Fe$^{2+}$ cation with the fully filled state in spin majority and only one electron occupied on $Y_{2}^{2}$ in the spin minority.
This result is demonstrated by Fig.\ref{fig:intro}(b). 
With the occupation and orbital components derived, one can estimate the SOC energy $\Delta E$ of $Y_2^2$ by $\Delta E=0.904\times\lambda (2\times 2\alpha + 1 \times 2\beta)\approx$ 32.9 meV, where $\lambda\approx 19$ meV is the SOC coefficient\cite{SOC_1962}. 
It matches well with the final PMA of 32.5 meV for Fe(1ML)/GaN. 
Therefore, PMA in Fe/GaN thin films are dominated by the first order perturbation of SOC.

According to the discussion above, large band splitting and the partial occupation in consequence are the precursor of large PMA. 
However, one should note that the SOC of Fe($3d$) is on the scale of 20 meV, two orders of magnitude smaller than the band width ($\sim 2.2$ eV) of $x^2-y^2/xy$ orbitals, labeled as $\Delta$ in Fig.\ref{fig:orbital}(f). 
SOC alone can hardly generate such a large splitting of the entire band. 
This is resonated by the fact that in simple non-self-consistent calculations, the PMA contributed by SOC alone is only 3.0 meV, a typical value in the second order correction scheme. 
Since the spin splitting changes slightly after SOC included and no structural reconstruction driven by SOC happens, SOC cannot be the major driving force of the band splitting. 
The only interaction on the scale of eV under investigation is the on-site electron-electron correlation interaction described by the Hubbard $U$. 
It thus suggests that the correlation interaction between occupied electrons must play a significant role on large PMA here. 

The energy contribution from Hubbard $U$ can be given by the single-particle expression under the Dudarev formation of L(S)DA$+U$ \cite{Dudarev1998}; $V_{m}^{\sigma}=(U-J)(1/2-n_{m}^{\sigma})$, where $U - J = 4.0$ eV is the $U$ value chosen for our first-principles calculations and $n_{m}^{\sigma}$ denotes the occupation number of orbital $m$ in spin channel $\sigma$. 
The energy of the $Y_{2}^{2}$-dominated state is $-1.62$ eV and that of the $Y_{2}^{-2}$-dominated state is $1.98$ eV, leading to a total splitting is $3.6$ eV. 
It is consistent with the splitting in PDOS [Fig.\ref{fig:orbital}(d)] and well exceeds the band width of the original $x^2-y^2/xy$ orbitals. 
On the other hand, no band splitting takes place if SOC is turned off since both $e$ and $e'$ states receive the same energy shift from the Hubbard $U$ and acquire the same occupation number due to the degeneracy. 
Therefore, the band splitting is triggered by SOC but amplified by the Hubbard $U$.

To further confirm this conclusion, we performed the SOC-included self-consistent calculations with multiple values of the Hubbard $U$. 
As shown in Fig.\ref{fig:orbital}(g), the splitting between $Y_{2}^{2}$- and $Y_{2}^{-2}$-dominated states is reduced when $U$ decreases. 
Fig.\ref{fig:orbital}(h) shows that PMA keeps a high value when $U = 2.0\sim5.0$ eV, and drops significantly when the Hubbard $U = 1.0$ eV, becoming smaller than the band width of the original $x^2-y^2/xy$ orbitals. 
Eventually, when the Hubbard $U$ is zero, almost equal populations of $ Y_{2}^{2}$ and $Y_{2}^{-2}$ states is recovered, and the magnitude of PMA is only 1.88 meV/u.c., entering the regime dominated by the second order correction of SOC. 
These results also confirm that the value of $U = 4.0$ eV is reasonable for our calculations since PMA is insensitive to $U$ when $U$ is higher than 2.0 eV.
Thus, the Hubbard $U$ plays a key role on the SOC-driven band splitting and the consequential large PMA in this thin film system.

\begin{figure*}
\includegraphics[scale=0.25]{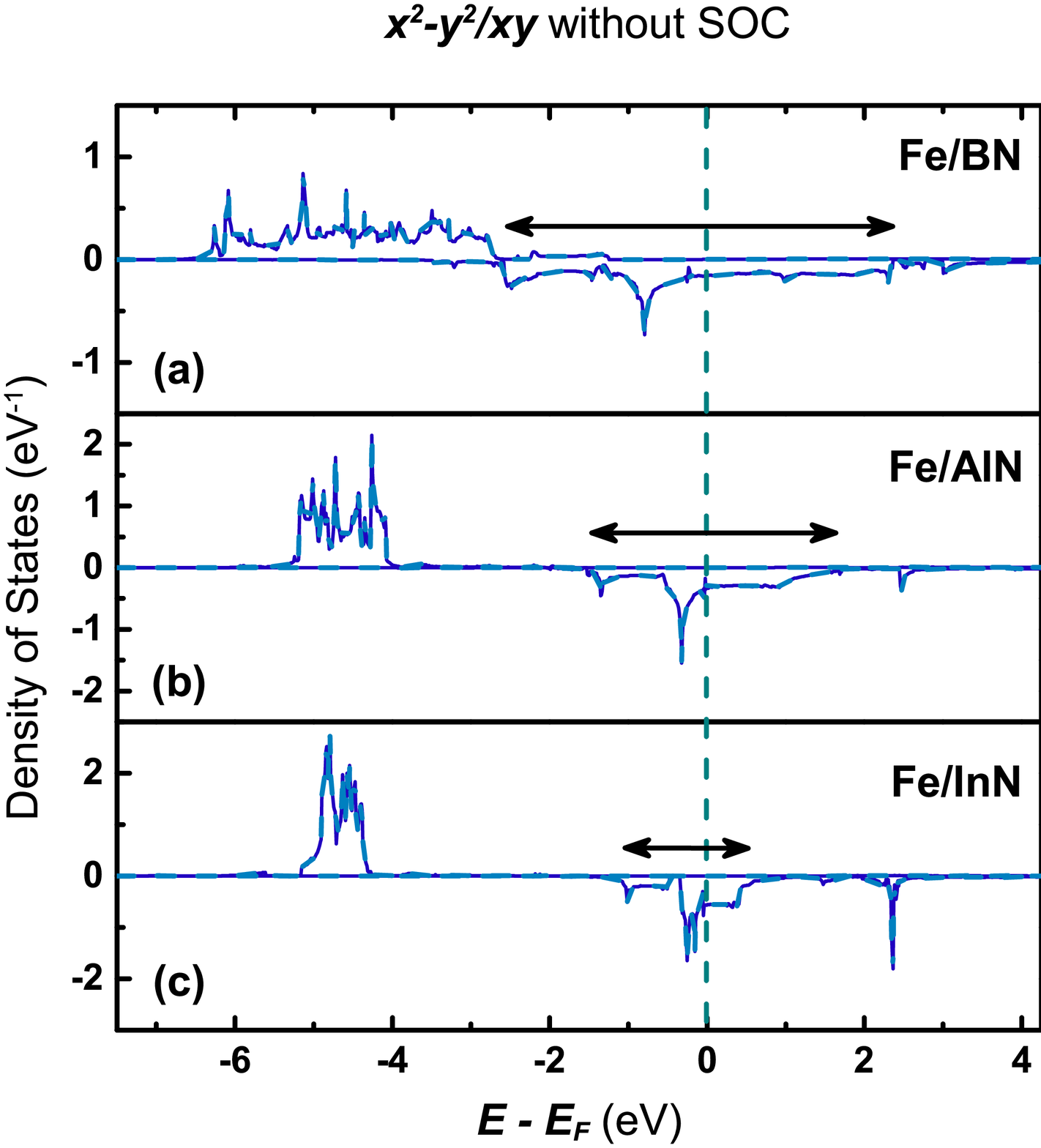}
\includegraphics[scale=0.25]{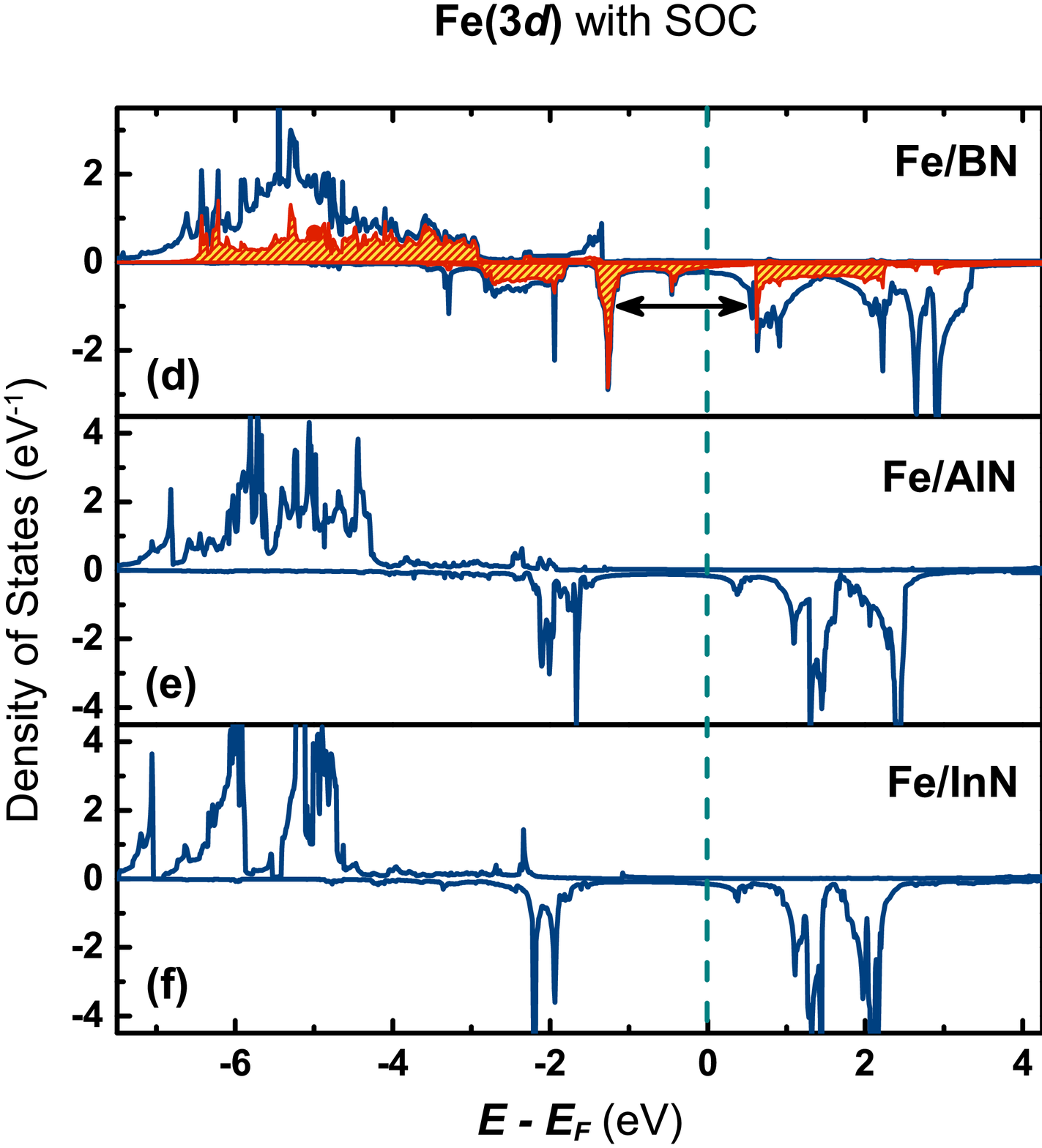}
\caption{
Electronic structures of 1ML Fe on BN, AlN and InN, respectively. 
(a)-(c) PDOS of $x^2-y^2/xy$ orbitals without SOC. 
The band width in spin minority channel is indicated by double arrows.
(d)-(f) PDOS of Fe($3d$) with SOC. 
In (d), the shadow region gives the projection onto $x^2-y^2/xy$ orbitals where the magnitude of splitting is indicated by the double arrow.
}
\label{fig:other} 
\end{figure*}


Electronic structures of 1 ML Fe on N-terminated (000$\bar{1}$) surface of BN, AlN and InN share the same behavior. 
The PDOS of Fe($3d$) $x^2-y^2/xy$ orbitals without SOC are shown in Fig. \ref{fig:other}(a)-(c).
Again the $x^2-y^2$ and $xy$ orbitals are degenerate due to the $C_3v$ symmetry. 
They are fully filled in the spin majority channel and partial filled in the spin minority one.
Although valence states of Fe cations in these structures are similar to that of Fe(1ML)/GaN, the band width shows strong dependence on the lattice constant of the substrate. 
It is $\sim$ 5.0, 3.0 and 1.5 eV for Fe/BN, Fe/AlN and Fe/InN respectively.
Therefore, Fe/BN with the smallest lattice constant has the largest band width due to large overlap of $x^2-y^2/xy$ orbitals lying in the plane. 
A band width of 5.0 eV there exceeds the magnitude of the Hubbard $U$ so that the combination of SOC+$U$ can hardly induce a large splitting of $x^2-y^2/xy$ orbitals into $Y_{2}^{2}$ and $Y_{2}^{-2}$.

It is confirmed by the PDOS of Fe($3d$) orbitals with SOC, as shown in Fig. \ref{fig:other}(d)-(f).
Relevant magnetic and orbital properties for each Fe(1ML)/III-V nitride is listed in Table \ref{table:pma}.
Considerable splitting around 3 eV at Fermi level of spin minority is found in Fe/AlN, Fe/GaN and Fe/InN, but is reduced to $\sim$1.5 eV in Fe/BN, as shown in Fig. \ref{fig:other}(e).
According to Table \ref{table:pma}, such small splitting in Fe/BN is consistent with the occupation number of the $Y_{2}^{2}$-dominated state, which is 0.724 for Fe/BN, smaller than 0.854 for AlN, 0.904 for GaN and 0.930 for InN.
Lifting of the degeneracy between $x^2-y^2$ and $xy$ orbitals gives a net orbital magnetic moment along $z$ direction.
It is 0.91 $\mu_{B}$ for Fe/BN, which is much smaller than 1.44 $\mu_{B}$ for Fe/AlN, 1.54 $\mu_{B}$ for Fe/GaN and 1.51 $\mu_{B}$ for Fe/InN. 
Spin moment for Fe/BN is 3.56 $\mu_B$. It is also the smallest among all III-V nitrides.
As a result, Fe/BN has the lowest PMA, at 24.1 meV/u.c., of all four materials under investigation. 
Still, it is an order of magnitude larger than any other transition metal thin films ever reported. 
On the other side, the large lattice constant of InN leads to small energy dispersion of the degenerate $x^2-y^2/xy$ orbitals and therefore results in a nearly full splitting between $Y_{2}^{2}$- and $Y_{2}^{-2}$-dominated states after SOC is turned on. 
A PMA of 53.7 meV/u.c for Fe/InN is consequently the largest, and almost hits the atomic limit of anisotropy energy of Fe.
Giant PMA for all four system follows the first order perturbation scheme of SOC.
The magnitude of PMA in units of mJ/m$^{2}$ is also listed in Table \ref{table:pma}.
In these units, PMA in Fe/BN is no longer the smallest due to small unit cell size in BN.

We further investigated the thickness dependence of PMA by using Fe/GaN $(000\bar{1})$ system as an example. 
Slab supercells with 2ML and 3ML Fe cations on top of GaN were built following the hexagonal closed packing along wurtzite GaN$(000\bar{1})$ direction. 
As a result, PMA for Fe(2ML)/GaN and Fe(3ML)/GaN are 37 meV/u.c.(58.4 mJ/m$^{2}$) and 21 meV/u.c.(33.2 mJ/m$^{2}$) respectively, which are still large enough to be considered in the first order SOC perturbation regime.

Fe thin films grown on the ($000\bar{1}$) N-terminated surface of III-V nitrides XN can be formed experimentally by the adlayer enhanced lateral diffusion method \cite{GaN_2003, InN_2009}.
As the X-terminated surface is found to be the most stable structure, at least for $1 \times 1$ ($000\bar{1}$) surface of GaN and InN \cite{GaN_1997, InN_2007}, one monolayer of Fe is grown on X-terminated $000\bar{1}$ surface first.
Then the N atoms from N$_2$ or NH$_3$ plasma are deposited into the space between Fe adlayer and top X layer, and gradually diffuse laterally under Fe adlayer.
Finally, a whole ML of N can be formed between Fe and X layers, resulting in a good surface morphology of  Fe(1ML)/III-V nitrides.

\subsection*{CONCLUSION}
In summary, $(000\bar{1})$ surface of III-V nitrides provide a crystal field of $C_{3v}$ symmetry. 
Using non-collinear spin-polarized first-principles calculations, we discovered a giant PMA in the 1ML Fe thin film on this $(000\bar{1})$ N-terminated surface of III-V nitride substrate. 
PMA ranged from 24.1 in BN to 53.7 meV/u.c. in InN substrate.
They are exceedingly large compared to existing PMA thin films and approach to the atomic limit of SOC energy of Fe. 
Electronic structure calculations and ligand field analysis show that each Fe cation has a net orbital angular momentum $L_z = 2$ originated from the symmetry-guaranteed splitting of $x^2-y^2$ and $xy$ orbitals. 
The on-site correlation interaction amplifies the splitting so that PMA is dominated by first order perturbation which is linearly proportional to the strength of SOC of Fe. 
Thickness dependence shows that PMA keeps a large value in multiple Fe layers deposited on GaN. 
It eases the experimental realization of our theoretical prediction.

In the rapidly developing technology of MRAM, lack of large PMA becomes a bottleneck in down sizing the binary bits. 
Giant PMA discovered here suggests that a 2.0 nm $\times$ 2.0 nm flake of Fe(1ML)/III-V nitride has a total uniaxial magnetic anisotropy energy about 1.2 eV, reaching the criteria for 10-year data retention at room temperature\cite{Dieny2017}.
Therefore, giant PMA in this thin film can ultimately lead to nanomagnetism and promote revolutionary ultra-high storage density in the future.
Furthermore, Large anisotropy energy could lead to large coercivity.
Fe/III-V nitride could lead to a new type of permanent magnet without rare earth element potentially.

\subsection*{METHODS}
The calculations were carried out in the framework of the non-collinear spin-polarized first-principles calculations with the projector augmented wave (PAW) pseudopotential \cite{PAW_1994} 
implemented in the Vienna ab initio simulation package (VASP) \cite{Kresse_1996_CMS}.  
We employed the generalized gradient approximation (GGA) of Perdew-Burke-Ernzerhof (PBE) formation \cite{PBE} plus Hubbard $U$ (GGA$+U$) \cite{Dudarev1998} with $U = 4.0$ eV on Fe($3d$) orbitals.

To build the slab supercell, four X-N (X = B, Al, Ga, In) principal layers are used as the substrate, and one to three Fe monolayers (MLs) are deposited on the N-terminated $(000\bar{1})$ surface. 
Dangling bonds of X-terminated $(0001)$ surface on bottom of the substrate are saturated by pseudo-hydrogen atoms. 
Vacuum layers in the supercells are more than 14 \AA\@ thick and the lattice constant in the plane is fixed to be 2.554 \AA, 3.113 \AA, 3.183 \AA\@ and 3.456 \AA\@ for BN, AlN, GaN and InN respectively, determined from the GGA-PBE result of the wurtzite phase of III-V nitrides. 
Positions of X and N atoms are borrowed from their bulk values. 
Positions of Fe atoms are optimized under collinear magnetic calculations without SOC until the force on each Fe atom is less than 1 meV/\AA. 
It is confirmed that N-top are the lowest energy sites of Fe atoms, as shown in Fig.\ref{fig:intro}(a).

Charge density of the SOC-free ground state was used as the initial state. Self-consistent total energy calculations were employed to derive the non-collinear calculation with SOC included. $\Gamma$-centered $25\times 25\times 1$ K-point meshes in the two-dimensional Brillouin zone were used with an energy cutoff of 600 eV for the plane-wave expansion. The accuracy of the total energy is thus guaranteed to be better than 0.1 meV per unit cell (u.c.).

\begin{figure}
\includegraphics[scale=0.30]{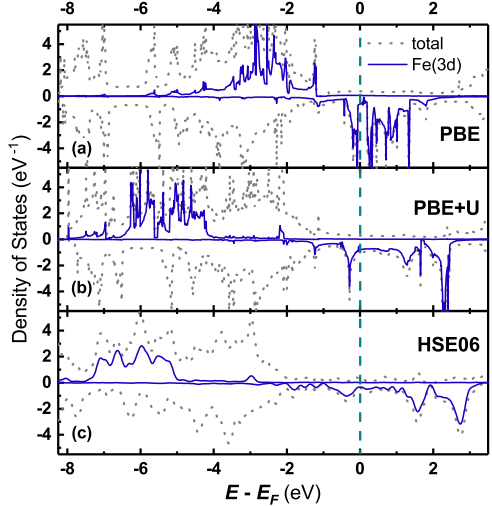}
\caption{DOS (gray dashed line) and PDOS (solid blue line) of Fe(1ML)/GaN without SOC derived by (a) GGA-PBE, (b) GGA$+U$ at $U=4.0$ eV and (c) HSE06 respectively. 
Consistency between (b) and (c) shows that GGA$+U$ at $U=4.0$ eV can well describe Fe($3d$) orbitals in this system.} 
\label{fig:u} 
\end{figure}


To obtain a reasonable $U$ value for GGA$+U$ calculations, we compared density-of-state (DOS) and projected density-of-state (PDOS) of 1 ML Fe on GaN$(000\bar{1})$ [Fe(1ML)/GaN] without SOC included by GGA$+U$ at $U = 4.0$ eV with those by GGA and the Heyd, Scuseria, and Ernzerhof hybrid functional (HSE06) 
\cite{HSE06} 
respectively.
HSE06 well describes the occupied states in magnetic systems by hybridizing the exact Fock exchange energy with GGA. To integrate over occupied states, tetrahedron smearing was used in GGA and GGA$+U$, while the Gaussian smearing with sigma value 0.05 eV was used in HSE06. According to the PDOS shown in Fig. \ref{fig:u}, spin majority channels of Fe (3d) for GGA, GGA$+U$ at $U = 4.0$ eV and HSE06 were mainly located at -4.0 $\sim$ -1.0 eV, -7.0 $\sim$ -4.0 eV and -7.5 $\sim$ -4.5 eV, respectively. Meanwhile, three peaks of the spin minority channel are located at -0.4 eV, 1.6 eV and 2.7 eV in HSE06, and at -0.3 eV, 1.3 eV and 2.3 eV in GGA$+U$, quite consistent to each other. As a comparison, all spin minority states are hybridized near the Fermi level from pure GGA calculations, completely different as the HSE06 result. Therefore, we used GGA$+U$ ($U = 4.0$ eV) for all calculations excpet for special notes.

\subsection*{Acknowledgements}
\textbf{Funding}:This work was supported by the U.S. Department of Energy (DOE), Office of Science, Basic Energy Sciences (BES) under Award No. DE-SC0016424 and used the Extreme Science and Engineering Discovery Environment (XSEDE) under Grant No. TG-PHY170023.  

\textbf{Competing interests}: JX.Y. and J.Z. are inventors on a U.S. provisional patent application related to this work filed by the the University of New Hampshire (application no. 62/589,399.; filed 21 November 2017)

\textbf{Author Contributions}: JX.Y. and J.Z. conceived the project together. JX.Y. carried out the first-principles calculations. Both authors discussed the results and wrote the manuscript.


\end{document}